\documentclass[twocolumn,aps]{revtex4}

\newcommand {\bea}{\begin{eqnarray}}
\newcommand {\eea}{\end{eqnarray}}
\newcommand {\be}{\begin{equation}}
\newcommand {\ee}{\end{equation}}

\usepackage{graphicx}

\begin{document}

\title{Meson Supercurrent State in High Density QCD}

\author{T.~Sch\"afer}

\affiliation{
Department of Physics\\
North Carolina State University,
Raleigh, NC 27695}

\begin{abstract}
The ground state of three flavor quark matter at asymptotically
large density is believed to be the color-flavor-locked (CFL) phase.
At non-asymptotic density the effect of the non-zero strange quark 
mass cannot be neglected. If the strange quark mass exceeds
$m_s\sim m_u^{1/3}\Delta^{2/3}$ the CFL state becomes unstable 
toward the formation of a neutral kaon condensate. Recently, several 
authors discovered that for $m_s\sim (2\Delta p_F)^{1/2}$ the CFL 
state contains gapless fermions, and that the gapless modes lead to 
an instability in current-current correlation functions. Using an 
effective theory of the CFL state we demonstrate that this instability 
can be resolved by the formation of a meson supercurrent, analogous to 
Migdal's $p$-wave pion condensate. This state has a non-zero meson 
current which is canceled by a backflow of gapless fermions. 

\end{abstract}
\maketitle


 Calculations based on weak coupling QCD show that the ground state of 
three flavor QCD at very high baryon density is the color-flavor-locked 
(CFL) phase \cite{Alford:1999mk,Schafer:1999fe,Evans:1999at}. During the 
past several years a lot of effort has been devoted to the question how 
the CFL state is modified in the presence of a non-zero strange quark mass. 
Using a BCS-type gap equation it is easy to see that the most important 
parameter is $m_s^2/(p_F\Delta)$, where $p_F$ is the Fermi momentum and 
$\Delta$ is the gap in the chiral limit \cite{Alford:1999pa}.

  Using effective field theory methods we showed that the CFL state 
is not modified if $m_s< m_s^{\rm crit}\sim m_u^{1/3}\Delta^{2/3}$.
At $m_s({\rm crit})$ the CFL phase undergoes a phase transition to 
a kaon condensed phase \cite{Bedaque:2001je}. This transition is shifted 
to larger values of $m_s$ if instanton effects are important, as is the 
case at moderate density \cite{Schafer:2002ty}. For $m_s^2/(2p_F)\sim 
2\Delta$ the effective field theory description breaks down and we expect 
a phase transition to a less symmetric phase to occur. More recently, 
several groups observed that gapless fermion modes can appear in the 
spectrum of the CFL phase in the vicinity of the point $m_s^2/(2p_F)\sim 
\Delta$ \cite{Alford:2003fq,Ruster:2004eg}. Gapless modes also appear in 
the kaon condensed CFL phase, but the critical strange quark mass is 
shifted to somewhat larger values \cite{Kryjevski:2004jw}.

 The problem is that gapless fermion modes in a weakly coupled (BCS) 
superfluid tend to cause instabilities in current-current correlation 
functions. These instabilities are quite generic and appear in a wide 
range of systems, including the 2SC phase, the CFL phase, and cold 
atomic gases \cite{Wu:2003,Huang:2004bg,Casalbuoni:2004tb}. The presence 
of an instability implies that the homogeneous superfluid is not the 
correct ground state \cite{Reddy:2004my,Giannakis:2004pf,Casalbuoni:2005zp}.
One possibility is that the true ground state is an inhomogeneous 
superconductor of the type first considered by Larkin, Ovchinnikov, 
Fulde and Ferrell \cite{Larkin:1964} and generalized to QCD in 
\cite{Alford:2000ze,Casalbuoni:2005zp}. In this work we wish to 
study the possibility that the ground state has a non-zero meson
current similar to the $p$-wave meson condensate suggested by Migdal, 
Sawyer and Scalapino \cite{Migdal:1973}. For this purpose we shall 
study the stability of the $s$-wave kaon condensate with respect to 
the formation of a non-zero current. Our study is analogous to the 
calculations performed in chiral models of nuclear matter 
\cite{Baym:1979} and in effective theories of cold atoms \cite{Son:2005qx}.

  Our starting point is the effective theory of the CFL phase
derived in \cite{Casalbuoni:1999wu,Kryjevski:2004jw}
\bea 
\label{l_cfl}
{\cal L} &=&  \frac{f_\pi^2}{4} {\rm Tr}\left(
  \nabla_0 \Sigma \nabla_0 \Sigma^{\dagger} - v_\pi^2 
  \vec{\nabla} \Sigma \vec{\nabla}\Sigma \right) \nonumber \\
& & \mbox{}
 + a_3 \left( \left({\rm Tr}\left[ M \Sigma\right]\right)^2 -
  {\rm Tr}\left[ M \Sigma M \Sigma  \right] + h.c. \right) \nonumber \\
 & & \mbox{}+ 
 {\rm Tr}\left(N^\dagger iv^\mu D_\mu N\right)  
 - D\ {\rm Tr} \left(N^\dagger v^\mu\gamma_5 
               \left\{ {\cal A}_\mu,N\right\}\right) \nonumber \\
 & & \mbox{}
 - F\ {\rm Tr} \left(N^\dagger v^\mu\gamma_5 
               \left[ {\cal A}_\mu,N\right]\right)
  \nonumber \\
 & &  \mbox{} + \frac{\Delta}{2} \left\{ 
     \left( {\rm Tr}\left(N N \right) 
   - \left[ {\rm Tr}\left(N \right)\right]^2 \right)  
     + h. c.  \right\}.
\eea
The effective lagrangian contains Goldstone boson fields $\Sigma$ and 
baryon fields $N$. The meson fields arise from chiral symmetry breaking 
in the CFL phase \cite{Alford:1999mk} and the baryon fields originate 
from quark-hadron complementarity \cite{Schafer:1999ef}. The chiral field 
is given by $\Sigma=\exp(i\phi^a\lambda^a/f_\pi)$ where $f_\pi$ is the 
pion decay constant. $M$ is the quark mass matrix. The chiral field and 
the mass matrix transform as $\Sigma\to L\Sigma R^\dagger$ and  $M\to 
LMR^\dagger$ under chiral transformations $(L,R)\in SU(3)_L\times 
SU(3)_R$. The baryon field $N$ transforms according to the adjoint 
representation of flavor $SU(3)$. We will expand $N$ in terms of the 
baryon nonet $(p,n,\Sigma^\pm, \Xi^\pm,\Xi^0,\Lambda^{8,0})$. $v^\mu
=(1,\vec{v})$ is the Fermi velocity, and $\Delta$ is the superfluid 
gap. We have suppressed the singlet fields associated with the breaking 
of the exact $U(1)_V$ and approximate $U(1)_A$ symmetries.

 The covariant derivative of the nucleon field is given by $D_\mu N=
\partial_\mu N +i[{\cal V}_\mu,N]$. The vector and axial-vector currents 
are 
\be
 {\cal V}_\mu = -\frac{i}{2}\left\{ 
  \xi \partial_\mu\xi^\dagger +  \xi^\dagger \partial_\mu \xi 
  \right\}, \hspace{0.2cm}
{\cal A}_\mu = -\frac{i}{2} \xi\left(\partial_\mu 
    \Sigma^\dagger\right) \xi , 
\ee
where $\xi$ is defined by $\xi^2=\Sigma$. It follows that $\xi$ transforms 
as $\xi\to L\xi U(x)^\dagger=U(x)\xi R^\dagger$ with $U(x)\in SU(3)_V$. 

 Symmetry arguments can be used to determine the leading mass terms in 
the effective lagrangian. Bedaque and Sch\"afer observed that $X_L=
MM^\dagger/(2p_F)$ and $X_R=M^\dagger M/(2p_F)$ enter the microscopic 
theory like the temporal components of left and right handed flavor 
gauge fields. We can make the effective lagrangian invariant under 
this symmetry by introducing the covariant derivatives 
\bea
\label{V_X}
 D_0N &=& \partial_0 N+i[\Gamma_0,N],  \\
 \nabla_0\Sigma &=& \partial_0\Sigma+iX_L\Sigma-i\Sigma X_R
\eea
with $\Gamma_0 = -\frac{i}{2}\{ \xi (\partial_0+ iX_R)\xi^\dagger 
+ \xi^\dagger (\partial_0+iX_L) \xi \}$. If the density is very large 
and the QCD coupling is weak then the low energy constants can be 
calculated in perturbation theory. The leading terms in the meson 
and baryon sector were calculated in \cite{Son:1999cm} and 
\cite{Kryjevski:2004jw}. The results are
\bea 
\label{f_pi}
f_\pi^2=\frac{21-8\log(2)}{18}
  \left(\frac{p_F^2}{2\pi^2}\right), \hspace{0.25cm}
v_\pi^2= \frac{1}{3}, \hspace{0.25cm}
a_3 = \frac{3\Delta^2}{4\pi^2},
\eea
and $D=F=1/2$. The gap $\Delta$ can also be computed in perturbative 
QCD, but we will not make use of these results and treat $\Delta$ as
a parameter.  

 The parameter that controls the effect of the strange quark mass
is the effective chemical potential $\mu_s=m_s^2/(2p_F)$. When $\mu_s$ 
exceeds the mass of the kaon the CFL ground state becomes unstable and 
a Bose condensate of kaons is formed. If isospin was an exact symmetry 
the energy the $K^+$ and $K^0$ meson would be degenerate. Because
of explicit isospin breaking, and because of the constraint of 
electric charge neutrality, a condensate of neutral kaons is favored. 
The $K^0$ condensed phase is characterized by  
\be 
\xi_{K^0} = \left( \begin{array}{ccc}
 1 &      0          &     0            \\
 0 & \cos(\alpha/2)  & i \sin(\alpha/2) \\
 0 & i\sin(\alpha/2) & \cos(\alpha/2) 
\end{array} \right),
\ee
with $\cos(\alpha)=m_K^2/\mu_s^2$. From equ.~(\ref{f_pi}) we 
get $m_K^2 = 3m_u(m_d+m_s)\Delta^2/(\pi^2 f_\pi^2)$. For typical
values of the parameters $\cos(\alpha)\sim (m_u/m_s)(\Delta/f_\pi)^2
\ll 1$. In the following we shall assume that $\cos(\alpha)=0$ 
and the ground state is a maximal kaon condensate. 

 Properties of the kaon condensed state were studied in 
\cite{Bedaque:2001je,Kryjevski:2004jw}. We showed, in particular, that 
the kaon condensed CFL phase has a gapless fermion mode that appears at 
$\mu_s=4\Delta/3$. For comparison, in the ordinary CFL phase gapless modes 
appear at $\mu_s=\Delta$ \cite{Alford:2003fq}. Recently, several groups 
have shown that gapless fermion modes lead to instabilities in the 
current-current correlation function \cite{Huang:2004bg,Casalbuoni:2004tb}. 
Motivated by these results we examine the stability of the kaon condensed 
phase against the formation of a non-zero current. 

 Consider a spatially varying $U(1)_Y$ rotation of the maximal 
kaon condensate
\be 
U(x)\xi_{K^0} U^\dagger (x) = \left(
 \begin{array}{ccc}
 1 & 0 & 0 \\
 0 & 1/\sqrt{2} & ie^{i\phi_K(x)}/\sqrt{2} \\
 0 & ie^{-i\phi_K(x)}/\sqrt{2} & 1/\sqrt{2} 
\end{array} \right).
\ee
This state is characterized by non-zero currents
\bea 
\label{cur}
\vec{\cal V} &=&  \frac{1}{2}\left(\vec{\nabla} \phi_K\right) \left(
 \begin{array}{ccc}
0 & 0 & 0 \\
0 & 1 & 0 \\
0 & 0 & -1 
\end{array} \right),\\
\vec{\cal A} &=&  \frac{1}{2}\left(\vec{\nabla} \phi_K\right) \left(
 \begin{array}{ccc}
0 & 0 & 0 \\
0 & 0 & -ie^{i\phi_K} \\
0 & ie^{-i\phi_K} & 0 
\end{array} \right).
\eea
In the following we compute the vacuum energy as a function 
of the kaon current $\vec{\jmath}_K=\vec\nabla\phi_K$. The meson 
part of the effective lagrangian gives a positive contribution
\be
{\cal E}=\frac{1}{2}v_\pi^2f_\pi^2\vec{\jmath}_K^2 .
\ee
A negative contribution 
can arise from gapless fermions. In order to determine this
contribution we have to calculate the fermion spectrum in 
the presence of a non-zero current. The relevant part of the
effective lagrangian is  
\bea 
{\cal L} &=& {\rm Tr}\left(N^\dagger iv^\mu D_\mu N\right)
 + {\rm Tr}\left(N^\dagger \gamma_5 \left( \rho_A+\vec{v}\cdot
      \vec{\cal A}\right) N\right) \nonumber \\
 & & \mbox{}  
 +\frac{\Delta}{2} \left\{ {\rm Tr}\left(N N\right) -
  {\rm Tr}\left(N\right){\rm Tr}\left(N\right)+ h.c.\right\},
\eea
where we have used $D=F=1/2$. The covariant derivative is 
$D_0N=\partial_0N+i[\rho_V,N]$ and $D_iN=\partial_i N +i
\vec{v}\cdot[\vec{\cal V},N]$ with $\vec{\cal V},\vec{\cal A}$ given in 
equ.~(\ref{cur}) and  
\be 
\rho_{V,A} = \frac{1}{2}\left\{ 
  \xi \frac{M^\dagger M}{2p_F}\xi^\dagger \pm 
  \xi^\dagger \frac{MM^\dagger}{2p_F} \xi 
  \right\}.
\ee
The vector potential $\rho_V$ and the vector current $\vec{\cal V}$
are diagonal in flavor space while the axial potential $\rho_A$ 
and the axial current $\vec{\cal A}$ lead to mixing. In order to 
exhibit the mechanism for the instability we will first discuss
the simple case of vanishing axial vector couplings $F=D=0$. We 
will then study the realistic case $D=F=1/2$. 

 We can determine the spectrum by decomposing $\rho_V$ and
$\vec{\cal V}$ into isospin and hypercharge components. We find
\be
\rho_V = -\frac{\mu_s}{2}\hat{I}_3
         -\frac{\mu_s}{4}\hat{Y},\hspace{0.5cm}
\vec{\cal V}= -\frac{\vec{\jmath}_K}{2} \hat{I}_3
              +\frac{3\vec{\jmath}_K}{4}\hat{Y},
\ee
where $\hat{I}_3=\lambda_3/2$ and $\hat{Y}=\lambda_8/\sqrt{3}$ are
the isospin and hypercharge generators. The dispersion relations 
can now be obtained from the $(I_3,Y)$ quantum numbers of the 
baryon nonet. The result is
\bea
 \omega_{n\ ,\Xi^0} &=& \sqrt{\Delta^2+l^2} 
  \pm  \vec{v}\cdot\vec{\jmath}_K, \nonumber \\
\label{spec}
 \omega_{\Sigma^+,\Sigma^-} &=& \sqrt{\Delta^2+l^2} 
  \pm \frac{1}{2}\left(\mu_s + \vec{v}\cdot\vec{\jmath}_K\right), \\
 \omega_{p\ ,\Xi^-} &=& \sqrt{\Delta^2+l^2} 
  \pm \frac{1}{2}\left(\mu_s - \vec{v}\cdot\vec{\jmath}_K\right).
\nonumber 
\eea
where $l=\vec{v}\cdot\vec{p}-p_F$ is the residual momentum. 
The neutral modes $(\Sigma_0,\Lambda_0,\Lambda_8)$ are not 
affected by the current or the vector potential. Equ.~(\ref{spec})
shows that there are gapless fermion modes as $\mu_s$ 
approaches $2\Delta$, and that gapless modes occur earlier if 
there is a non-zero current. The contribution to the vacuum 
energy from gapless modes is 
\be
\label{e_fct} 
{\cal E} = 4 \frac{\mu^2}{2\pi^2}\int dl \int 
 \frac{d\Omega}{4\pi} \;\omega_l \theta(-\omega_l) ,
\ee
where the factor 4 is a degeneracy factor and $d\Omega$ is an 
integral over the Fermi surface. Near the Fermi surface we can 
approximate
\be
\omega_l = \Delta +\frac{l^2}{2\Delta}-\frac{1}{2}
 \left(\mu_s +\vec{v}\cdot\vec{\jmath}_K\right).
\ee
The integral in equ.~(\ref{e_fct}) receives contributions from 
one of the pole caps on the Fermi surface. 
The result has exactly the same structure as the energy 
functional of a non-relativistic two-component Fermi 
liquid with non-zero polarization which was recently analyzed
by Son and Stephanov \cite{Son:2005qx}. Introducing dimensionless 
variables 
\be 
 x = \frac{j_k}{a\Delta}, \hspace{0.5cm}
 h = \frac{\mu_s-2\Delta}{a\Delta}.
\ee
we can write ${\cal E} = c {\cal N} f_h(x)$ with
\be
 f_h(x) = x^2-\frac{1}{x}\left[
   (h+x)^{5/2}\Theta(h+x) - (h-x)^{5/2}\Theta(h-x) \right] .
\ee
We have defined the constants
\be
\label{consts}
 c = \frac{8^4}{2\cdot 15^4c_\pi^3 v_\pi^6},\hspace{0.25cm}
 {\cal N} =   \frac{\mu^2\Delta^2}{\pi^2},\hspace{0.25cm}
 a = \frac{8^2}{15^2 c_\pi^2 v_\pi^4}, 
\ee
where $c_\pi = (21-8\log(2))/36$ is the numerical coefficient 
that appears in the weak coupling result for $f_\pi$. 
According to the analysis in \cite{Son:2005qx} the function
$f_h(x)$ develops a non-trivial minimum if $h_1<h<h_2$ with 
$h_1\simeq -0.067$ and $h_2\simeq 0.502$. In perturbation 
theory we find $a=13.9$ and the kaon condensed ground state
becomes unstable for $(\Delta- \mu_s/2) < 0.46\Delta$. 

\begin{figure}[t]
\includegraphics[width=8.75cm]{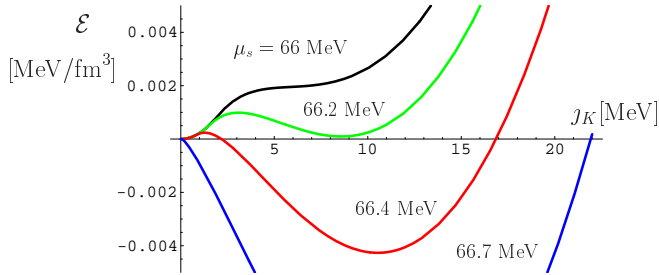}
\caption{Energy density as a function of the current $\jmath_K$
for several different values of $\mu_s=m_s^2/(2p_F)$ close
to the phase transition.}
\label{fig_jfct}
\end{figure}

 We will see that the instability is weaker if the axial coupling 
is taken into account. The axial vector potential and the axial vector 
current can be written as 
\be
 \rho_A +\vec{v}\cdot\vec{\cal A} = 
   \frac{i}{2}\left (\mu_s+\vec{v}\cdot \vec{\jmath}_K \right)
  \left(e^{i\phi_K}\hat{u}^+ - e^{-i\phi_K}\hat{u}^-\right),
\ee
where $\hat{u}^\pm = (\lambda_6\pm i \lambda_7)/2$ are the $u$-spin
raising and lowering operators. The axial vector coupling leads
to mixing among the charged $(p,\Xi^-,\Sigma^\pm)$ and neutral 
$(n,\Xi^0,\Sigma^0,\Lambda^{0,8})$ baryons. Gapless fermions occur 
in the charged sector. The mixing in the charged sector is  
\be 
\label{l_mix}
{\cal L } = \frac{i}{2} 
   \left( \mu_s+\vec{v}\cdot\vec{\jmath}_k \right)
   \left\{ e^{i\phi_K} \left(\overline{\Xi}^-\Sigma^-\right)
         -e^{-i\phi_K}\left(\overline{\Sigma}^-\Xi^-\right) \right\} .
\ee
The dispersion relation of the lowest mode is given by
\be
\label{disp_ax}
\omega_l = \Delta +\frac{(l-l_0)^2}{2\Delta}-\frac{3}{4}
  \mu_s -\frac{1}{4}\vec{v}\cdot\vec{\jmath}_K,
\ee
where $l_0=(\mu_s+\vec{v}\cdot\vec{\jmath}_K)/4$. The calculation
of the energy density is completely analogous to the one we 
performed in the previous section. The only difference is that 
the number of gapless modes is reduced, and that the effect of 
the current on the energy of the fermion is smaller. The energy 
functional is ${\cal E} = c{\cal N}f_h(x)$ with $x$ defined in 
eq.~(\ref{consts}) and $h = (3\mu_s-4\Delta)/(a\Delta)$. The
numerical coefficients are $c = 2/(15^4c_\pi^3 v_\pi^6)$ amd 
$a= 2/(15^2 c_\pi^2 v_\pi^4)$. 

Using the weak coupling values for $v_\pi$ and $c_\pi$ we find 
$a=0.43$ and the kaon condensed ground state becomes unstable 
for $(\Delta- 3\mu_s/4) < 0.007\Delta$. In Fig.~\ref{fig_jfct} 
we show the energy density as a function of the current $\jmath_K$ 
for several values of $\mu_s$ near the transition.We note that the 
mechanism for the instability is quite robust, but the value of the 
onset chemical potential depends sensitively on the values of the low 
energy constants. We also note that the homogeneous gapless kaon 
condensed phase becomes stable at even larger values of the effective 
chemical potential, $(\Delta-3\mu_s/4)<-0.05$. This result may not be 
reliable, however, because the contribution from higher fermion modes 
has to be included.  
 
 We note that the ground state has no net current. This is clear 
from the fact that the ground state satisfies $\delta {\cal E}/\delta 
(\vec\nabla\phi_K)=0$. As a consequence the meson current is canceled 
by an equal but opposite contribution from gapless fermions. This is 
analogous to what happens in a $p$-wave pion condensate in nuclear  
matter \cite{Baym:1973,Baym:1979,Migdal:1973}
or in the LOFF phase \cite{Alford:2000ze}. We also expect that the 
ground state has no chromo-magnetic instabilities. The effective 
lagrangian given in equ.~(\ref{l_cfl}) is formulated in terms of 
gauge invariant fields. Quark-hadron continuity implies that we
should study the screening length of a $SU(3)_F$ gauge field. 
For diagonal gauge fields we find
\be
m_V^2=\left. \frac{\partial^2\mathcal E}{\partial \jmath_K^2}
  \right|_{\jmath_K =0} 
 = v_\pi^2 f_\pi^2\left(1 - \frac{5}{8\sqrt{h}}\theta(h)\right),
\ee
which shows the magnetic instability for $h>0$ and has the characteristic 
square root singularity observed in microscopic calculations 
\cite{Huang:2004bg}. In this work we have calculated the energy functional 
for arbitrary values of $\jmath_K$, not just the second derivative at the 
origin. We find that the phase transition to a supercurrent state is first 
order and that the instability sets in for $h>h_c$ with $h_c<0$. We also 
observe that the second derivative at the new minimum is positive and of 
the same order of magnitude as the screening mass at $h=0$.  

 Summary: We have that shown that for $\mu_s\sim 4\Delta/3$, which is 
the threshold for the appearance of gapless fermion modes, the CFL 
phase becomes unstable toward the formation of a meson supercurrent.
This state is analogous to the $p$-wave pion condensate studied in 
low density nuclear matter. In that context the name $p$-wave was 
motivated by the fact that the interaction that leads to the formation 
of a condensate is the $v\cdot D$ interaction between nucleons and pions 
that also determines the $p$-wave $\pi N$ scattering amplitude. This is 
also true in our case. Supercurrent formation is driven by a $v\cdot D$ 
interaction between baryons and mesons. 

 Our result suggests that the series of phases encountered in three 
flavor QCD as the density is lowered starting from an asymptotically large 
value consists of the CFL phase, an $s$-wave kaon condensate, and a $p$-wave 
kaon condensate. There are many questions that remain to be addressed. In 
this work we established the presence of an instability but did not determine 
the exact nature of the new ground state. We have to consider additional 
currents, the role of other light fermions, and the effects of electric
charge neutrality.  Once the currents become large the effective field 
theory description breaks down and more microscopic approaches along the 
lines of \cite{Buballa:2004sx,Forbes:2004ww} are necessary. These methods 
might also be useful in order to determine the fate of the meson 
supercurrent state as the density is lowered even further and $\mu_s$ 
approaches $2\Delta$. One possibility is that the supercurrent phase 
evolves into a LOFF-type state. 

Acknowledgments: I would like to thank A.~Kryjevski 
for showing me a draft version of his closely related work 
\cite{Kryjevski:2005qq}. This work is supported in part by the US 
DOE grant DE-FG-88ER40388.


\end{document}